# ラムダアーキテクチャによるリアルタイム予測保守

山登庸次† 熊崎宏樹† 福本佳史†

† NTT ソフトウェアイノベーションセンタ
東京都武蔵野市緑町 3-9-11
E-mail: †{yamato.yoji,kumazaki.hiroki,fukumoto.yoshifumi}@lab.ntt.co.jp

**あらまし** 近年，IoT 技術が進展しており，メンテナンス分野での応用が期待されている．しかし，現場状況をリアルタイムに分析できていない，データ収集のコストが高い，故障検知ルールの設定コストが高い等の課題から，日本では十分広がっていない．本稿は，これら課題解決のため，ラムダアーキテクチャの考えを用いて，エッジはセンサデータを解析しリアルタイムにアノマリーを検知し，新たなルールを抽出するとともに，クラウドはバッチで収集されたデータを分析しエッジの学習モデルをより高い精度に更新する，メンテナンスプラットフォームを提案する，更に，サンプルアプリケーションを実装する．

# Realtime Predictive Maintenance with Lambda Architecture

Yoji YAMATO†, Hiroki KUMAZAKI†, and Yoshifumi FUKUMOTO†

† Software Innovation Center, NTT Corporation
3-9-11, Midori-cho, Musashino-shi, Tokyo 1808585 Japan
E-mail: †{yamato.yoji,kumazaki.hiroki,fukumoto.yoshifumi}@lab.ntt.co.jp

**Abstract** Recently, IoT technologies have been progressed and applications of maintenance area are expected. However, IoT maintenance applications are not spread in Japan yet because of insufficient analysis of real time situation, high cost to collect sensing data and to configure failure detection rules. In this paper, using lambda architecture concept, we propose a maintenance platform in which edge nodes analyze sensing data, detect anomaly, extract a new detection rule in real time and a cloud orders maintenance automatically, also analyzes whole data collected by batch process in detail, updates learning model of edge nodes to improve analysis accuracy.

## 1. はじめに

近年，IoT（Internet of Things）技術やクラウド技術（例えば [1]）が進展している．IoT の適用範囲は多岐に渡るが，その中でも，ドイツの Industrie 4.0 [3] 構想で進められている，製造やメンテナンス分野が適用先として有力視されている．工場，機械，製品等の情報をセンサにより収集，分析することで，それらの状態を可視化し，生産把握，計画反映，物流制御，不良物品交換等のサプライチェーンの自動化，サービス連携（例えば [2]）による，ビジネスの加速が期待されている．

IoT データを用いた，IoT アプリケーションを開発，運用するための，IoT プラットフォームも出てきている [4], [5]．しかし，既存の IoT プラットフォームは，大量のセンサデータを収集し可視化することが主な注力点となっており，前述のようなメンテナンスの加速に生かすことは十分されておらず，日本では十分普及していない．具体的には，現場状況をリアルタイムに分析できていない，データ収集のネットワークコストが高い，故障検知ルールの設定コストが高い問題が上げられる．

本稿では，工場での業務機器を題材に，これら課題を解決するメンテナンスプラットフォームを提案する．提案プラットフォームは，エッジはセンサデータを解析しリアルタイムにアノマリーを検知し新たなルールを抽出するとともに，クラウドはバッチで収集されたデータを分析しエッジの学習モデルをより高い精度に更新する．更に，提案プラットフォーム上に，サンプルアプリケーションを実装する．

## 2. 既存技術の課題

既存 IoT プラットフォーム技術の課題を整理する．

AWS IoT [4] は，Amazon Web Services の各種機能を IoT で統合利用可能にするプラットフォームである．例えば，Amazon

Kinesis を用いて，MQTT(MQ Telemetry Transport) プロトコルでデータを収集し，取集したデータを Amazon Machine Learning の機械学習機能により回帰やクラス分類の多彩な分析等が可能である．

NTT ドコモと GE（General Electric）社は，GE の産業用機器向けワイヤレスルーター Orbit と，ドコモの通信モジュールを連携させた IoT ソリューションを 2015 年に発表している [5]．企業は，遠隔地の設備にドコモの通信モジュールを内蔵した GE の Orbit を設置することで，設備の稼働データを収集できる．さらに，ドコモが提供する Toami 上で IoT アプリケーションを開発可能であり，取得データの可視化や Web サービス連携（ [6] [7] 等）を容易にしている．

しかし，これらの技術をメンテナンスに利用する際は，大きく 3 つ課題があると考える．

第一に現場状況をリアルタイムに分析できていない. [5] は，集約データのバッチ処理による可視化が中心であり，リアルタイムな分析に基づく部品手配等には繋がっていない．Industrie4.0 では，事前に定義した異常ルールに基づく流通手配等は想定されているが，工場の環境や季節等の条件で適切なルールが変わる現場状況に追従することはできていない．

第二にセンシングデータ収集のコストが高い．AWS IoT では分析のため一度全データをクラウドに集めるが，多くの機械のセンサデータを，各地から収集するためのネットワークが必要である．例えば，IoT によるメンテナンス事例として著名なコマツ社の KOMTRAX は，建設機械のデータ収集に衛星通信を使っており，コストが大きいと共に，全てのデータは送れずデータを間引いて送っている．

第三に故障検知のためのルール設定のコストが高い．様々な IoT センサデータから，故障を検出するためには，分析アプリケーション側でルールや閾値を設定し判定することが主流である．例えば，PSPP [8] 等の統計解析ソフトウェアを用いて，その判定ルールを抽出する方法があるが，ルール，閾値設定は高度な知識が必要であり，その設定コストは高い．

これらの課題から，日本ではメンテナンス分野に IoT が十分普及しているとは言えない．

## 3. ラムダアーキテクチャコンセプトに基づくメンテナンスプラットフォームの提案

前節の課題を解決するため，lambda アーキテクチャのコンセプトを用いたメンテナンスプラットフォームを提案する．ラムダアーキテクチャは，分析結果をユーザに提供する際に，バッチレイヤー，スピードレイヤーの両方で分析することで，細やかな集計結果と速報性の必要な結果を実現する，Marz が提案したアーキテクチャである. [9]

図 1 にラムダアーキテクチャの適用イメージを示す．工場等のエッジ側には，データ保持用のストレージに加え，Jubatus [10] が配置される．Jubatus は，ストリームデータの逐次処理に適した機械学習フレームワークである．クラウド側には，CEP (Complex Event Prosessing), DB, メンテナンスアプリケーションが配置される．

まず，スピードレイヤーでは，エッジ側の Jubatus が，センサデータをストリーム処理し，異常疑いのあるデータを検知し，その場合にクラウド側の CEP にデータを送信する．Jubatus は機械学習を通じて新たな異常疑いルールも抽出して送信する．CEP を介して異常疑いデータを受け取ったメンテナンスアプリケーションは，故障予測等の分析をし，必要に応じて外部システム（ERP（Enterprise Resource Planning）等）と連携する．外部システム連携は，Web サービス等の既存連携技術（例えば [11]）を用いればよい．次に，バッチレイヤーでは，エッジ側の生データが，夜間等のコストが安い時間に，クラウド側に送信され，DB に保持される．メンテナンスアプリケーションは，スピードレイヤーでは対象外の詳細な分析を，全データを用いて分析する．更に，データサイエンティストは，定期的に，生データを分析して，より精度高い機械学習モデルを抽出し，エッジ側に配信することもできる．エッジ側の分析精度が十分高くなった際には，バッチレイヤーは不要としてもよい．

図 1 のアイデアにより，前節課題が解決される．Jubatus によりリアルタイムで異常疑いを検知し，交換部品発注等の即時のアクションに繋げることができる．また，Jubatus は，通常のデータ値と異なる場合を検知するアノマリー検知（LOF（Local Outlier Factor）[12] 等）等のアルゴリズムが利用できるため，現場環境データに応じたルールや閾値を抽出し，クラウドに送信し，メンテナンスアプリケーションに反映できる．スピードレイヤーでは，異常疑いデータのみ送信することでネットワークコストを下げることができる．バッチレイヤーで必要な全データは，夜間や物理的輸送等の手段で，コストを下げることができる．

図 2 は，上記アイデアに基づいた提案プラットフォームのアーキテクチャである．図 2 を参照して，処理ステップを説明する．

1. エッジには，業務機器にセンサーが設置され，センシングデータが随時収集され保存される．エッジノードの Jubatus は，ストリーム処理でデータをアノマリー検知や分類分析をし，通常運用と大きく異なる場合に，異常を判定し，クラウド側に関連したデータを送信する．Jubatus はオンラインで判定だけでなく，学習もできるため，新たな異常疑いルールを抽出した場合は，それもクラウド側に送られる．2. エッジ側とクラウド側の送受信には，MQTT, HTTP 等のプロトコルが使われ，CEP を介してメンテナンスアプリケーションや DB へデータが受け渡される．3. メンテナンスアプリケーションは，異常疑いデータから PSPP 等の統計解析ソフトウェアを用いて，故障原因，時期等の故障予測を行う．4. 故障予測結果に基づいて，メンテナンスを手配する際は，ERP 等に交換部品の発注依頼がされる．

これらの通常メンテナンス業務とは独立に，エッジとクラウドの双方向でルールや学習モデルの更新がされる．まず，Jubatus により抽出された現場データに応じたルールは，クラウド側に送信され，PSPP 等の統計解析ソフトウェアで抽出されるルールにマージされる．これとは逆に，夜間等のバッチ処理でエッジ側の生データがクラウド側に収集される．これらデータはま

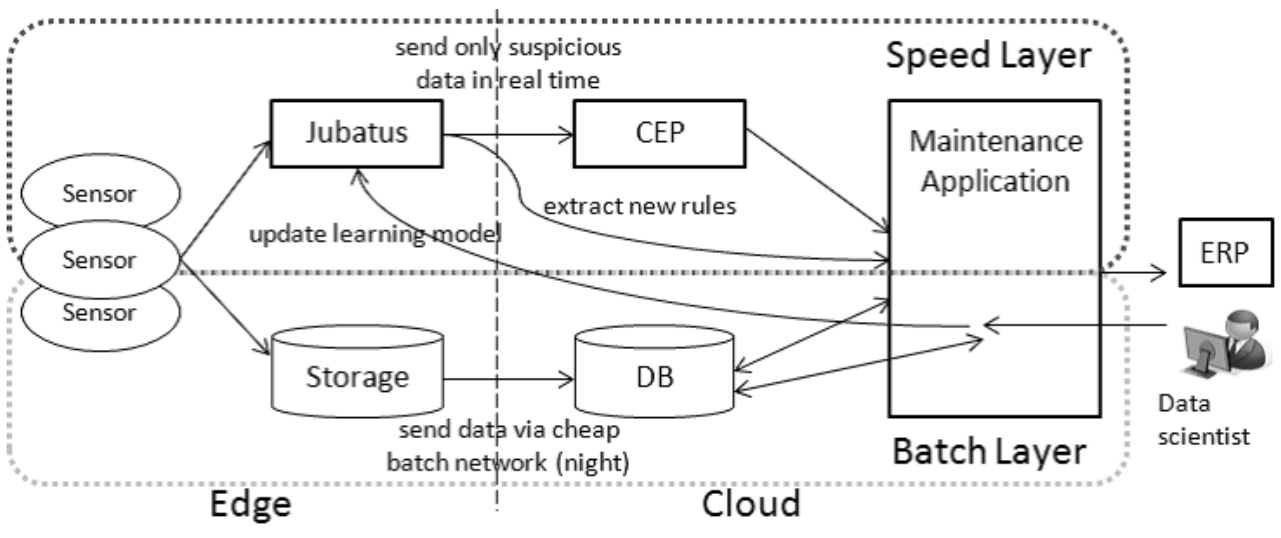

図 1　Lambda architecture adoption for maintenance analysis

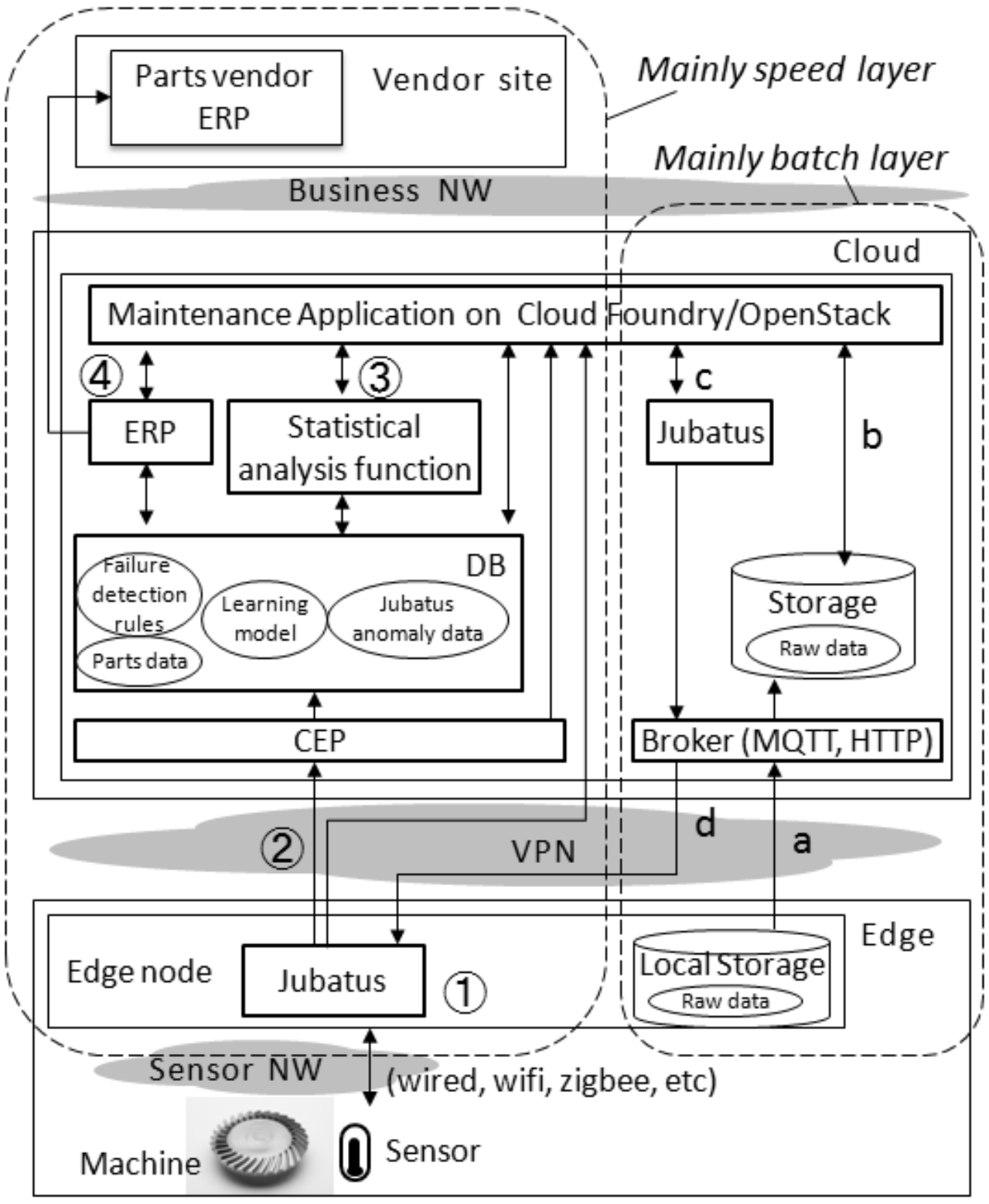

図 2　Proposed maintenance platform architecture

ず，前述のステップ 3 と同様に自動故障予測に用いられる．これは，速報性は不要だが詳細な予測をしたい場合ためである．データサイエンティストが定期的に，生データを分析して，クラウド側の Jubatus を用いてより精度高い検知ができる機械学習モデルを更新し，エッジ側に配信する．この双方向の更新により，クラウド側，エッジ側の分析精度を向上することが出来る．なお，設定更新時はクラウド更新技術 [13] や自動検証技術 [14] を用いればよい．

## 4. サンプルアプリケーション実装

前節のアーキテクチャに基づいて，サンプルアプリケーションを実装した．これは，日本国内の工場の業務機器のファンにマイクが付けられており，そのファンの異音から故障を予測し，メンテナンスに繋げるアプリケーションである．

図 3(a) の右は，工場のエッジを模擬している．エッジノードは，Raspberry Pi [15] で Jubatus0.8.1 がインストールされている．ファンの音は，マイクにより収集され，Raspberry Pi の Jubatus はその周波数データを逐次分析する．図 3(a) の右の

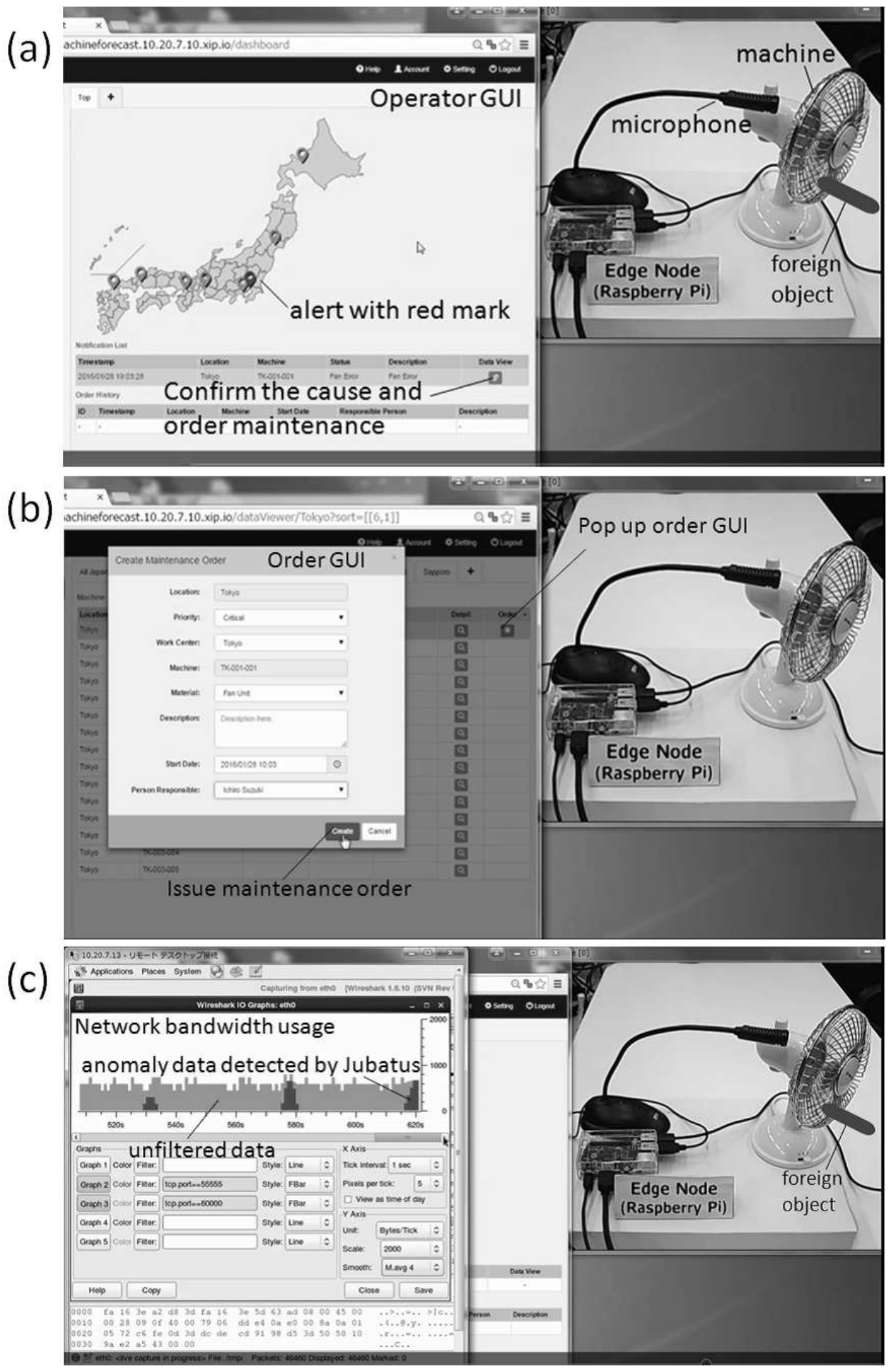

図 3　Implemented example application. (a) failure detection and shows an alert. (b) maintenance order creation to ERP. (c) comparing network bandwidth usage of unfiltered data and anomaly data detected by Jubatus.

ファンに異物を挿入すると異音が生じるため，Jubatus は異常を判定する．Raspberry Pi は，判定情報を含む異常データを，クラウド側に送信する．クラウド側では，異常データから故障を予測し，図 3(a) の左のように赤印でアラートを表示する．工場のオペレータは，アラートをクリックすることで，故障原因や部位が表示され，交換部品発注等の，メンテナンスオーダーが提案される（図 3(b)）．オペレータが提案を了解すると，オーダーが ERP と人材管理機能に送られる．

Jubatus のフィルタ効果の確認のため，図 3(c) は，ネットワーク利用帯域を示している．図 3(c) の左のグラフの内，フィルタされていないデータは緑で，Jubatus で検出された異常データは赤で表示されている．異物を図 3(c) 右のファンに入れると，異常が検出され関連データがクラウドに送信される．図 3(c) の左から，Jubatus がネットワーク帯域利用量を低減できていることが分かる．

## 5. 結　　論

本稿では，業務機械のメンテナンスを支援するプラットフォームを提案し，サンプルアプリケーションを実装をした．提案プラットフォームは，エッジで Jubatus によりストリームデー

タを分析し，異常疑いデータや新たな抽出ルールをスピードレイヤーとしてクラウドに送信し，クラウドは部品発注等のリアルタイムアクションに繋げるとともに，Jubatusの抽出ルールをクラウド側にマージしたり，バッチレイヤーの分析結果をJubatusの学習モデルに反映させ，クラウドやエッジの分析精度を向上させる．今後は，サンプルアプリケーションを用いて，実際の工場現場に提案し，検証，改善していく予定である．

## 文　献